# Optimal Engagement of Residential Battery Storage to Alleviate Grid Upgrades Caused by EVs and Solar Systems

Rafi Zahedi[*,1], Amirhossein Ahmadian[1], Chen Zhang[1], Shashank Narayana Gowda[1], Kourosh SedghiSigarchi[2], Rajit Gadh[1]

[1] Smart Grid Energy Research Center, University of California, Los Angeles, Los Angeles, 90095, USA

[2] ECE Department, California State University, Northridge, Northridge, 91330, USA

A B S T R A C T

The integration of distributed energy resources has ushered in a host of complex challenges, significantly impacting power quality in distribution networks. This work studies these challenges, exploring issues such as voltage fluctuations and escalating power losses caused by the integration of solar systems and electric vehicle (EV) chargers. We present a robust methodology focused on mitigating voltage deviations and power losses, emphasizing the allocation of a Permitted Percentage (PP) of battery-based solar systems within residential areas endowed with storage capabilities.

A key facet of this research lies in its adaptability to the changing landscape of electric transportation. With the rapid increase of electric trucks on the horizon, our proposed model gains relevance. By tactically deploying PP to oversee the charging and discharging of batteries within residential solar systems, utilities are poised not only to assist with grid resilience but also to cater to the upcoming demands spurred by the advent of new EVs, notably trucks.

To validate the efficacy of our proposed model, rigorous simulations were conducted using the IEEE 33-bus distribution network as a designed testbed. Leveraging advanced Particle Swarm Optimization techniques, we have deciphered the optimal charging and discharging commands issued by utilities to energy storage systems. The outcomes of these simulations help us understand the transformative potential of various PP allocations, shedding light on the balance between non-battery-based and battery-based solar residences. This research underscores the need for carefully crafted approaches in navigating the complexities of modern grid dynamics amid the anticipated increase in electric vehicles.

## 1. Introduction

High penetration of distributed energy resources (DER), such as solar systems could result in power quality issues on the distribution network [1,2]. This will be impacted more severely as the number of Electric Vehicles (EVs) grows [3].

Moreover, the emergence of electric trucks introduces distinct challenges and prospects for the distribution network. As highlighted in [4,5], if utilities neglect the increasing prevalence of electric trucks, the distribution network experiences notable voltage fluctuations. They suggested that with an increase in charging loads, there's a heightened probability for upgrades needed for the distribution lines offering an expensive and time-consuming solution with extensive planning to implement [4].

However, strategies for utilities to bear the costs remains an open-ended important question. Furthermore, a high penetration of Non-Battery-Based Solar Residence (NBBSR) elevates the likelihood of voltage fluctuations on distribution lines, particularly when a significant amount of solar power is injected into the feeders

In accordance with FERC Order 2222, regulatory bodies in the United States have been actively promoting the integration and adoption of various Distributed Energy Resources (DERs). This regulatory framework also facilitates the development of models for aggregated resource participation, enabling Behind-the-Meter (BTM) assets such as BTM Battery Energy Storage Systems (BESS) to provide a comprehensive range of power system services for which they possess the technical qualifications [6].

Moreover, in the context of bring-your-own-device business models, utility customers who are also energy producers, such as residential households, are increasingly investing in Battery-Based Solar Residence (BBSR) systems. These prosumers often utilize only a fraction of their storage capacity. In such scenarios, prosumers may consider sharing their unused storage capacity with third parties [7,8]. Additionally, through Energy-Storage-as-a-Service Arrangements, developers or utilities assume the initial costs associated with BBSR systems and subsequently own and manage these systems in return for a fee [9,10]. Both of these aforementioned business models enable utilities to expand the adoption of BBSR systems and exercise varying degrees of ownership and management as needed. For instance, New York ISO has designed a model that will let DERs such as BBSRs provide services to consumers, utilities, and the wholesale energy market to mainly ensure bulk power reliability and accessibility to all grid parties. However, they have not prioritized the power quality problems relating to the high penetration of NBBSR and EVs [11]. To mitigate the power quality challenges caused by DERs including EVs and solar systems, [12,13] provide more in-depth discussion. They suggested that utilizing storage systems could be potentially more cost-effective and merit more investigation for power quality and voltage deviation issues.

To underscore the importance of this study, a unique optimization approach is introduced and assessed within the context of the IEEE 33-bus system.

This approach aims to enhance the power quality of the grid, specifically focusing on mitigating active power losses and voltage deviations, all without the need for additional infrastructure upgrades such as storage systems, and the incorporation of new energy resources. To reduce the number of deciding factors without sacrificing its general applicability, the IEEE 33-bus system is divided into seven distinct sectors. Within each sector, there are multiple instances of two types of solar systems: BBSR and NBBSR. Notably, only NBBSR is authorized to supply solar energy to the grid, but a designated Permitted Percentage (PP) of the BBSR capacity is allocated to enable the utility to access it through charging and discharging commands as needed. Through exploration of various combinations of BBSR battery PPs and different ratios of NBBSR and BBSR at each bus, the study identifies the optimal charging and discharging commands from the grid for the existing storage capacity within each sector. This approach empowers utilities to determine the necessary PP levels by engaging in negotiations with BBSRs, taking into account the actual NBBSR and BBSR ratio and their desired power quality standards.

The remainder of this paper is categorized into four sections. In Section 2, details of the proposed methodology is investigated. A multi-objective optimization model is formulated in Section 3. The whole optimization flowchart is also introduced in this Section. The test case is analyzed in Section 4, and the results are demonstrated. Finally, the major contributions of the present work discussed in Section 5.

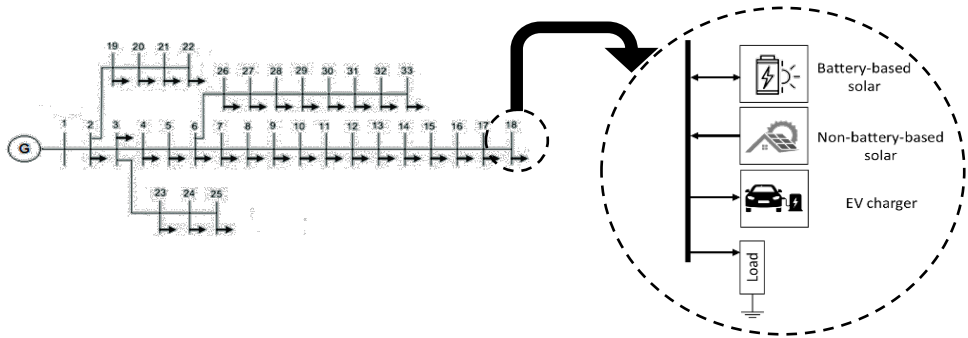

Figure 1: A typical structure of the Bus arrangement used in this study

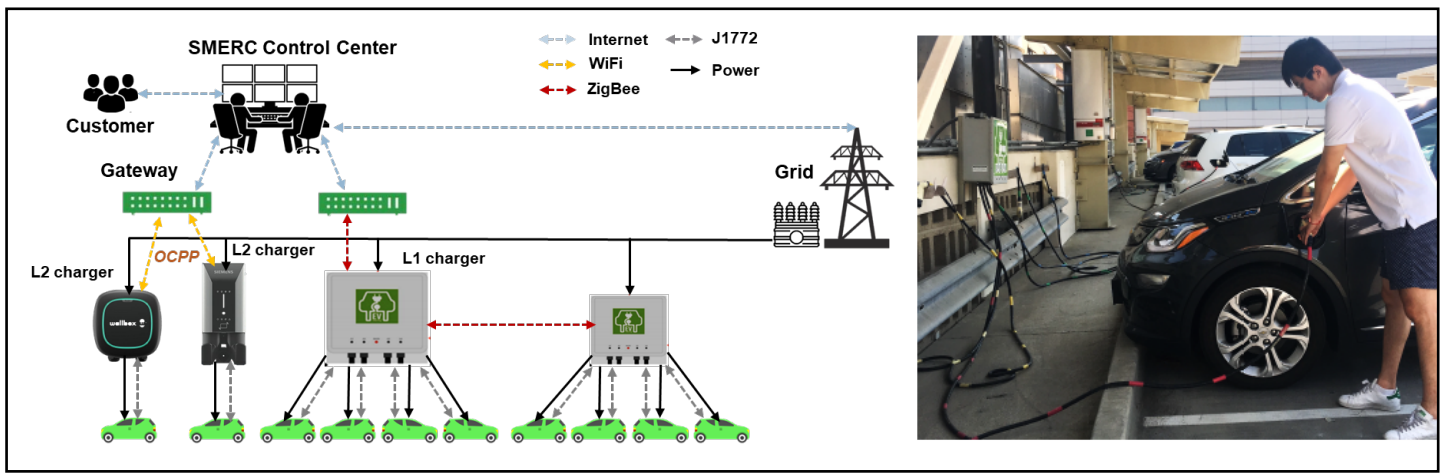

Figure 2: UCLA Smart Grid Energy Research Center charger network architecture

## 2. System Model

EV energy usage data was gathered from UCLA Smart Grid Energy Research Center charging station as illustrated in Figure 2. The parking structure has different level 1 and 2 chargers. The level 1 chargers talk among themselves and the gateway via ZigBee, and level 2 chargers talk to the gateway via OCPP.

The primary objective of this study is to address the challenges posed by the increasing adoption of EV chargers and solar systems within distribution networks, specifically targeting power losses and deviations in bus voltage. To achieve this goal, the proposed model has been developed in accordance with the IEEE-33 bus radial distribution standard based on [14]. It is assumed throughout this study that every residential dwelling connected to each bus incorporates a solar power generation system. Figure 1 presents an overview of the distribution system encompassing all buses adhering to the IEEE-33 [15] bus standard, encompassing EV chargers, two types of solar systems (BBSR and NBBSR), as well as residential load. Additionally, we introduce the concept of "PP," denoting a percentage of BBSR's capacity that utilities can employ to issue charging and discharging commands to/from the grid as needed.

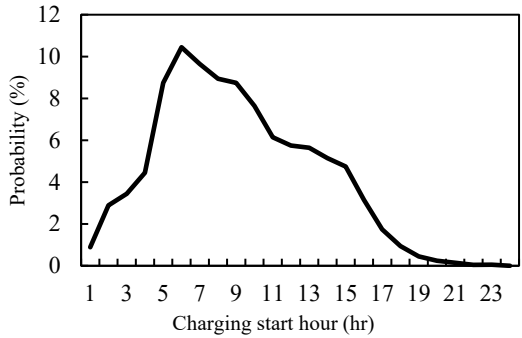

Figure 3: Distribution of the charging start hour for weekdays

Figure 3 shows our assumption on the average distribution of a charging start hour during weekdays, derived from [16].

Batteries serve as short-term energy storage devices and are commonly utilized in conjunction with solar systems, referred to as BBSR [17]. Furthermore, the energy available for each bus can be determined using the following calculation [17]:

$$E_{BT,max} = \beta \times N_{User} \times E_{BT,user} \qquad (1)$$

where $\beta$, $N_{User}$, and $E_{BT,user}$ represent the PP of BBSR's capacity, number of residential houses per each bus, nominal storage capacity of each BBSR (kWh), respectively. Additionally, for each time interval, the calculation of available energy per bus is as follows [17]:

$$E_{BT,i}(t) = E_{BT,i}(t-1) + \sum_{t=1}^{T} P_{BT(t)} \cdot \Delta(t) \qquad (2)$$

where T, $E_{BT,i}$ (kWh) and $P_{BT,i}$ (kW) are number of hours of operation, the stored energy, and the dispatched power from the utility determined by the optimization algorithm at time interval t for bus i. Finally, the state of charge of BBSR's batteries ($SOC_{BT}$) calculated as [17]:

$$SOC_{BT,i}(t) = 100 \times \frac{E_{BT,i}(t)}{E_{BT,max}} \qquad (3)$$

## 3. Proposed Methodology

The proposed approach is designed to reduce the total active power loss and voltage deviations over a 24-hour operational period, enabling the utility to implement optimal charging and discharging commands through Particle Swarm Optimization (PSO). Figure 4 illustrates the flowchart outlining this methodology [14].

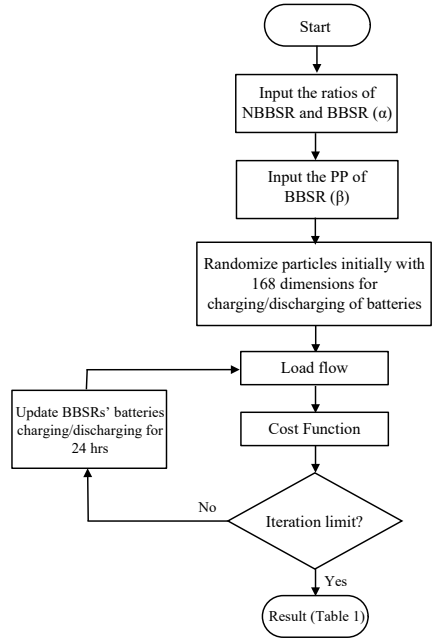

Figure 4: Flowchart of overall proposed optimization procedure

### 3.1. Cost Function

The cost function in this study can be expressed as the weighted combination of both active power loss and voltage deviation. Active power loss represents the amount of energy dissipated as heat in the distribution network, while voltage deviation measures the deviation of bus voltages from their nominal values. Voltage deviation exceeding a limit can affect electric equipment's lifetime and trigger the protection relays. They can be represented as follows [15]:

$$P_{loss} = \sum_{t=1}^{T} \sum_{k=1}^{N} I_{t,k}^2 R_{t,k} \qquad (4)$$

$$V_{dev} = \sum_{t=1}^{T} \sum_{k=1}^{N} |V_{rated} - V_{t,k}| \qquad (5)$$

$$\text{Cost Function} = \min \left( w_1 \cdot P_{loss} + w_2 \cdot V_{dev} \right) \qquad (6)$$

where $V_{rated}$, $V_K$, $V_{dev}$, $I_K$, $R_K$, $P_{Loss}$, T and N are rated voltage, the voltage of each bus, sum of all buses' voltage deviation, feeder current loading, line resistance, total active power loss, the total number of hours of operation and the total number of lines in the radial distribution network, respectively.

### 3.2. Particle Swarm Optimization

PSO is an optimization technique that was created by Kennedy and Eberhart [15]. It is inspired by the behavior of bird flocking and fish schooling. PSO is composed of a set of particles that make up a group. Each particle searches its local space to find the local minimum or maximum. The velocity and position of each particle can be updated according to its best experience and the best experience of its neighbors [15]. The variables of each PSO's particle are important factors to guarantee the optimal solution [16]. Using a revolutionary optimization algorithm for a high number of variables can entail significant drawbacks, including the potential for high computational costs and the curse of dimensionality, where the exponential growth in search space makes efficient exploration challenging. Additionally, understanding algorithm behavior and diagnosing issues can be harder, leading to overfitting, generalization problems, and difficulty in visualizing the optimization landscape. Finally, there's an increased risk of premature convergence to suboptimal solutions, making optimization in high-dimensional spaces particularly challenging.

The application of PSO in this study involves optimizing the charging and discharging commands of residential battery storage systems within the distribution network. PSO is employed to find the optimal values for parameters that govern the charging and discharging process, ensuring the reduction of active power losses and voltage deviations. Specifically, the PSO algorithm iteratively refines the charging and discharging commands based on the performance metrics defined by the cost function, converging towards an optimal solution that minimizes power losses and maintains grid voltage stability.

In distribution systems, the number of buses can be relatively high, which leads to weak optimization results. In this study, the IEEE-33 bus has been subdivided into seven sectors, as depicted in Figure 5, to effectively address the challenge posed by a high number of variables. This partitioning of the area is based on presumed demographic conditions and charging demands. Therefore, the charging/discharging command can be strategically allocated to each of these sectors to reduce the number of decision variables for the optimization algorithm (see Figure 5).

### 3.3. Load Flow

The backward/forward sweep method is used for load flow. The backward/forward sweep is an iterative method in which, at each iteration two computational stages are performed. It is one of the most effective methods for load flow of radial distribution systems [18]. The backward/forward sweep method was chosen for its effectiveness in analyzing radial distribution systems. This method allows for an iterative calculation of currents and voltages, starting from the load buses and moving towards the substation and vice versa. Its suitability for radial systems simplifies the load flow analysis, and its step-by-step approach facilitates understanding.

Below's a brief overview of how the backward/forward sweep load flow method works:

*Backward Sweep:*

- Start from the load buses (the farthest points from the substation) and work your way back toward the substation.
- At each load bus, calculate the current injected into the bus by the load.
- Use Kirchhoff's Current Law to calculate the current leaving the bus toward the substation.

*Forward Sweep:*

- Start from the substation and move outward to the load buses.
- At each bus (except the substation), calculate the current entering the bus based on the current calculated in the backward sweep.
- Use Ohm's Law to calculate the voltage at each bus based on the current and the impedance (resistance and reactance) of the transmission lines and transformers.

Active power consumption per bus is the variable for this method in each iteration, which is obtained from the following equations

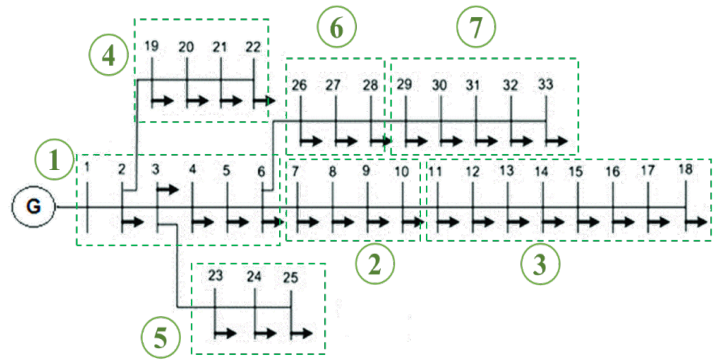

Figure 5: Modified IEEE-33 bus with 7 sectors

$$P_{Load,i}(t) = P_{T,Solar,i}(t) + P_{EV,i}(t) + P_{Res,i}(t) \qquad (7)$$

$$P_{T,Solar,i}(t) = \alpha \times N_{B,i} \times P_{Solar,i}(t) \qquad (8)$$

$$\alpha = \frac{N_{NBBSR}}{N_{NBBSR} + N_{BBSR}} \qquad (9)$$

where $P_{Load,i}$, $P_{T,Solar,i}$, $P_{Solar,i}$, $P_{EV,i}$, $P_{Res,i}$, $\alpha$, $N_{B,i}$, $N_{NBBSR}$, and $N_{BBSR}$ are load, total generated solar energy by NBBSR, generated solar per NBBSR unit, EV load, residential load, proportion coefficient, number of residences, number of NBBSR, and the number of BBSR at bus i, respectively.

## 4. Results

The IEEE-33 bus radial distribution system has been meticulously implemented as the test case using MATLAB 2022b, comprising 33 buses and an intricate network of 32 distribution lines. These lines exhibit varying current-carrying capabilities, with lines connecting node-1 to node-9 accommodating robust capacities of 400 A, while the remaining lines exhibit 200 A capacity. In this meticulously crafted system, the combined losses in active and reactive power are quantified at 281.58 kW and 187.95 kVAR, respectively. It's imperative to note that the baseline voltage is standardized at 11 KV.

The empirical foundation for this research derives from real-world data collected from the UCLA charging station, particularly focusing on EV charging behaviors. Furthermore, solar energy generation patterns for each NBBSR have been meticulously modeled, drawing inspiration from the empirical distributions captured in Figure 6. This data stems from the UCLA 35 kW solar plant, albeit downscaled to a maximum of 10 kWs to align with the research's temporal requirements. Complementing this, the EV load distribution per individual bus is aptly visualized in Figure 7, providing a granular perspective on energy consumption patterns.

The implementation of seven distinct control sectors is a pivotal aspect of this study. Remarkably, this strategic segmentation substantially reduces the number of optimization variables from an initial count of 768 (32 buses × 24 hours) down to 168 (7 sectors × 24 hours). It is essential to note that the number of residences for every bus is the same and equal to 92.

Through the deployment of the PSO algorithm, a comprehensive exploration of various combinations of the optimization parameters α and β is undertaken. The outcomes of this rigorous experimentation are shown in Figure 8, where it is observed that that an insightful pattern emerges. Voltage deviations from the nominal values and power losses exhibit decreasing trends with the progressive augmentation of β for each α. Moreover, the cost function showcases a dynamic response, reducing as α escalates from 0% to 80% and then increasing as α surpasses the 80%. Also, figure 8 indicates that the lowest cost function happens where α equals 70% and β equals 30%.

The numerical values for 77 evaluated combinations of α and β are summarized in Table 1. To show the significance of these findings, it should be noted that the cost function for an unoptimized network is 80. Notably, the achieved cost functions marked improvements in voltage stability and substantial reductions in power loss across the distribution network.

In alignment with the study's objectives, voltage limits within an acceptable range are strategically set at ±10% of the nominal voltage, as articulated in [19].

In radial distribution networks, the end bus of each branch usually faces the highest voltage drop. To evaluate the proposed methodology's efficacy, the voltage profiles of buses 18 and 33, which are the end bus of their own branches are analyzed. Figure 9 illustrates the voltage profiles of buses 18, representing the end buses of their branches. The profiles correspond to the configuration where α equals 70% and β equals 30%, achieving the lowest cost function as highlighted in Table 1. As seen in Figure 9, the proposed methodology consistently meets the specified voltage limits, showcasing its effectiveness in maintaining grid stability. While in other scenarios the voltage profile violates the 10% deviation. A similar pattern is observed in Figure 10 for the voltage profiles of bus 33, reinforcing the robustness of our approach in different scenarios.

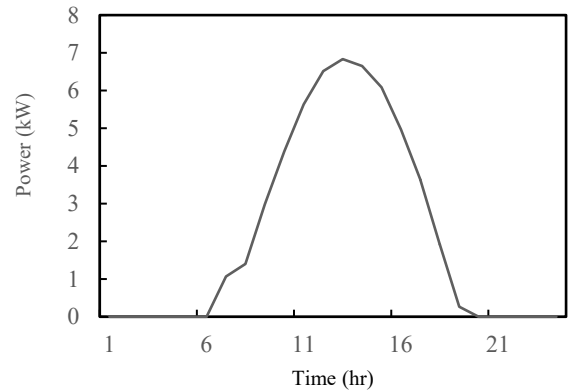

Figure 6: Generated solar

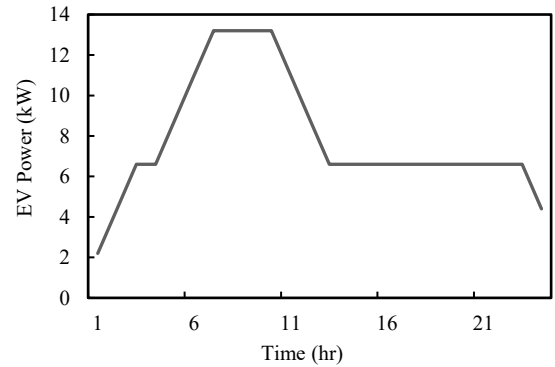

Figure 7: EV load per bus

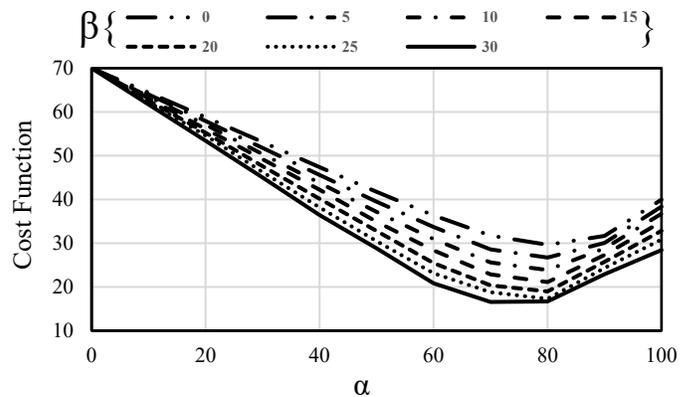

Figure 8: Cost Function value for different combination

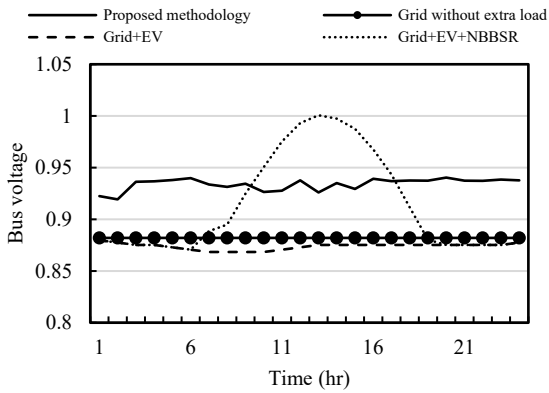

Figure 9: Bus 18 voltage for different scenarios during 24 hr of operation

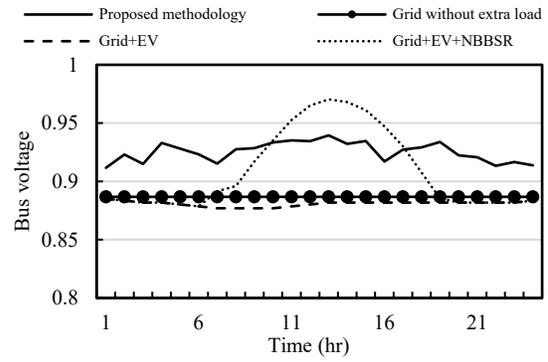

Figure 10: Bus 33 voltage for different scenarios during 24 hr of operation

Table 1 provides numerical values of Figure 8 for various combinations of α and β, showcasing the impact of these parameters on voltage stability, power losses, and the overall cost function. Table 2 further compares the proposed methodology with alternative scenarios. It provides a comprehensive comparative analysis among the proposed methodology with α set at 70% and β at 30% and alternative scenarios, considering key metrics such as total active power loss, average voltage deviation across all 33 buses, and cost function values.

The outcomes of our analysis reveal remarkable improvements in various performance metrics. Specifically, we observe a substantial 47.8% reduction in active power losses when compared to Grid without extra loads. Furthermore, the reduction in active power loss is a notable 53.2% compared with the Grid+EV scenario. Also, the proposed methodology registers a commendable 33.6% decrease in active power losses in comparison with GRID+EV+NBBSR scenario. These findings underscore the remarkable efficacy of our proposed methodology in enhancing the overall efficiency of the distribution network.

Table 1: Cost Function (equation 6) for different combination of α and β

| | | 0% | 5% | 10% | 15% | 20% | 25% | 30% |
|---|---|---|---|---|---|---|---|---|
| | 0% | 70.01663 | 70.01663 | 70.01663 | 70.01663 | 70.01663 | 70.01663 | 70.01663 |
| | 10% | 64.44564 | 63.98924 | 63.53546 | 63.09185 | 62.60636 | 62.22017 | 61.70375 |
| | 20% | 58.81144 | 57.91499 | 57.07909 | 56.09525 | 55.22481 | 54.61636 | 53.40737 |
| | 30% | 53.11391 | 51.76321 | 50.36399 | 49.20472 | 47.59238 | 46.2408 | 45.02529 |
| | 40% | 47.39317 | 45.58555 | 43.77159 | 42.20241 | 40.14026 | 38.16705 | 36.43257 |
| α | 50% | 41.75065 | 39.50142 | 37.14781 | 34.90428 | 32.78993 | 30.47759 | 28.8417 |
| | 60% | 36.36633 | 33.71627 | 30.93126 | 28.34218 | 25.45428 | 23.10252 | 20.82519 |
| | 70% | 31.73868 | 28.55686 | 25.57219 | 22.87464 | 20.36441 | 18.78012 | **16.55407** |
| | 80% | 29.62956 | 26.71179 | 23.85455 | 21.09941 | 18.96617 | 17.28412 | 16.66144 |
| | 90% | 31.67723 | 30.07172 | 28.833 | 27.30397 | 25.7509 | 24.32898 | 22.90161 |
| | 100% | 39.96917 | 38.34332 | 36.70755 | 34.89518 | 32.83336 | 30.8411 | 28.44189 |

Table 2: Comparison between the proposed methodology and other scenarios for 24 hr of operation

| | Grid without extra load | Grid+EV | GRID+EV+NBBSR | Proposed methodology |
|---|---|---|---|---|
| **Active power loss (kW)** | 6758.1 | 7541.6 | 5314.7 | 3527.4 |
| **Average voltage deviation (%)** | 6.99 | 7.41 | 5.33 | 4.34 |
| **Cost function** | 26.5094 | 28.6454 | 21.1694 | 16.55407 |

Our analysis demonstrates an impressive average voltage deviation reduction of 37.9% when contrasted with the Grid without extra loads. The Grid+EV scenario also benefits significantly, with a 41.4% reduction in average voltage deviation. Also, the proposed methodology can achieve 18.6% reduction in average voltage deviation of GRID+EV+NBBSR scenario. These results highlight the transformative potential of our approach in achieving enhanced grid performance and reliability across various scenarios.

The results are striking, revealing a substantial 37.55% reduction in the cost function when compared to Grid without extra loads, a remarkable 42.21% decrease relative to the Grid+EV scenario, and a noteworthy 21.8% deduction in comparison with the GRID+EV+NBBSR scenario.

## 5. Conclusion

In this study, we have presented an innovative approach to enhance grid power quality without extensive infrastructure upgrades. Our focus was on mitigating active power losses and reducing voltage deviations, critical for ensuring a reliable and stable distribution network. Utilizing the widely recognized IEEE 33-bus system as our testing ground, we explored various scenarios involving PP of BBSR and different ratios of NBBSR to BBSR at each bus. Through optimization, we determined efficient charging and discharging commands, minimizing the need for costly modifications.

Our research demonstrated that several combinations effectively met voltage drop limitations, highlighting the potential for utilities to proactively address challenges posed by DERs. Additionally, electric trucks, while environmentally promising, pose unique charging infrastructure challenges. Our methodology offers a solution by optimizing charging commands, reducing strain on local grids, minimizing disruptions, enhancing electric truck operations, and ensuring overall grid stability. This adaptable approach stands as a strategic tool for utilities and fleet operators, fostering efficiency and sustainability in the face of evolving energy demands.

The outcome of preventing the unnecessary upgrade of the grid from this study not only minimizes power losses and stabilizes the grid but also plays a pivotal role in reducing carbon emissions, particularly in regions where the energy grid relies on a mix of fossil fuels. Implementing the proposed methodology encourages both BBSRs and NBBSRs to expand their solar capacity. The promotion of renewable energy integration, coupled with the optimized utilization of resources, underscores the study's commitment to sustainable practices.

For future works, a notable suggestion involves analyzing the integration of heavy-duty electric vehicles into the grid and exploring the feasibility of utilizing distributed solar generation to meet their charging requirements. This direction addresses the evolving landscape of electric transportation, focusing on the unique challenges posed by heavy-duty electric vehicles and proposing a sustainable approach through the utilization of solar energy.

## Acknowledgment

The EV database used in the current research was created by way of a project sponsored in part by LADWP/DOE grant numbers 20699 and 20686, which were part of the LADWP Smart Grid Regional Demonstration Project awarded to the department of Mechanical and Aerospace Engineering and the Smart Grid Energy Research Center (SMERC) at UCLA. We are also grateful for the additional sponsorship received from the following grants 69763, 77739, and 45779 to the Department of Mechanical and Aerospace Engineering.

   7